\input{aipcheck}

\documentclass[
    ,final            
  ]
  {aipproc}

\layoutstyle{6x9}
\usepackage{amssymb}
\usepackage{aas_macros}

\begin{document}

\title{ Optical spectroscopy of (candidate) ultra-compact X-ray binaries}



\author{Gijs Nelemans}{
  address={Institute of Astronomy, Cambridge, UK \\and\\
Department of Astrophysics, Radboud University Nijmegen, The Netherlands}
}

\author{Peter Jonker}{
  address={Harvard-Smithsonian Center for Astrophysics, Cambridge, MA,
  USA}
}

\begin{abstract}
We present (preliminary) results of our systematic spectroscopic study
of (candidate) ultra-compact X-ray binaries. Most candidates are
confirmed and we found the first optical spectra of (pure)
carbon-oxygen accretion discs.
\end{abstract}

\maketitle


\section{Introduction}

Ultra-compact X-ray binaries (UCXBs) are close binaries with periods
less than about one hour, in which a neutron star (or possibly a black
hole) accretes material from a companion star. Their short periods
rule out ordinary hydrogen-rich companion stars, since these stars are
too big; they do not fit in the Roche lobe \citep[e.g.][]{nrj86}

In recent years a renewed interest has developed in these systems for
a number of reasons. Improved observing facilities (e.g. large optical
telescopes and sensitive X-ray satellites) make it possible to study
them in more detail. But more importantly the number of known
systems has increased. In particular the discovery of three transient
UCXBs \citep[e.g.][]{mss+02,rss02,mjs03} in which the millisecond
pulsations of the accreting neutron star were seen, has been an
exciting development.

We have started a systematic spectroscopic study of known and
candidate UCXBs (excluding systems in globular clusters) in order to
confirm/reject the candidates, to constrain possible formation
scenarios and to open the way to study the chemically peculiar
accretion process that is expected to operate in these binaries.

\section{Current open questions}

The formation of UCXBs has been proposed through three channels, in
which the donor stars are either white dwarfs, helium stars, or the
remnants of evolved main sequence stars \citep[see][and references
therein]{prp02}. However, there are still many open questions about
UCXBs. One of the first priorities is to increase the number of known
systems and to find their orbital periods. Besides the orbital period
there are a number of properties that indicate that a particular
system could be an UCXB.  \Citet{vm94} studied the absolute magnitudes
of X-ray binaries and derived a relation between the absolute
magnitude, the orbital period and the X-ray luminosity, based on the
assumption that the absolute magnitude is dominated by the irradiated
accretion disc, whose surface is determined by the size of the binary,
i.e. the orbital period. They empirically gauged this
relation. Systems with absolute magnitudes fainter than about $M_V =
4$ have a good chance to be UCXBs.

\begin{table}
\begin{tabular}{llllc}
\hline \hline Name & Period & Ne & M$_{\rm V}$& confirmed? \\
 & (min) & & & \\ \hline \hline 
4U 1543-624 & 18$^a$  & y & & $\checkmark$ \\ 
4U 0614+09 & ?  & y & 5.4 & $\checkmark$ \\ 
2S 0918-549 & ?  & y & 6.9 & ($\checkmark$) \\ 
4U 1822-00 & ?  & & 5.5 &  ($\checkmark$) \\ 
4U 1556-605 & ? & y & & $\times$\\ 
XB 1905+000 & ? & & 4.9 & ??\\  \hline \hline
$a$ \citet{wc04}
\end{tabular}
\caption{Overview of the UCXB candidates.}
\label{tab:candidates}
\end{table}

\citet{jpc00} found features in the X-ray spectrum of the 20 min UCXB
4U 1850-087, which they attributed to enhanced Ne in the system and
found similar features in 3 other systems, making them good UCXB
candidates. In Table~\ref{tab:candidates} we list our selection of
candidates, based either on their absolute magnitudes or the the
precence of the ``Ne feature''.

Optical spectroscopy of UCXBs might also be a good way to study the
formation of UCXBs, as the different formation scenarios will lead to
different chemical composition of the transferred material and thus of
the accretion disc. The first constraints on the chemical composition
have come from the properties of the type I X-ray bursts observed from
the 11 min globular cluster sources 4U 1820-30, suggesting helium rich
material \citep{bil95}.  In the following section we will describe the
results of our observations, which are already summarized in the last
column of the table.

\section{Optical spectroscopy: results}

\subsection{The VLT and FORS2}

We used the FORS2 spectrograph on the 8.2m Very Large Telescope (VLT)
of the European Southern Observatory at Paranal to obtain optical
spectra of our candidate UCXBs. The observations were taken in the
spring of 2003 and 2004. In 2003 we used the 1400V and 600RI
holographic grisms, with a 1" slit, using 2x2 on-chip binning.  This
setup resulted in coverage of 4620 -- 5930 \AA\ with mean dispersion
of 0.64 \AA/pix for the 1400V spectra and 5290 -- 8620 \AA\ with mean
dispersion of 1.63 \AA/pix for the 600RI spectra. The 2004 spectra are
taken with the 600B and 600RI grisms. The 600B grism covers the range
3325 -- 6367\AA\ with dispersion of 1.48 \AA/pix. All spectra were
reduced using standard IRAF tasks.

\subsection{4U 0614+09, 4U 1543-624 and 2S 0918-549}

\begin{figure}
  \includegraphics[angle=-90,width=\textwidth]{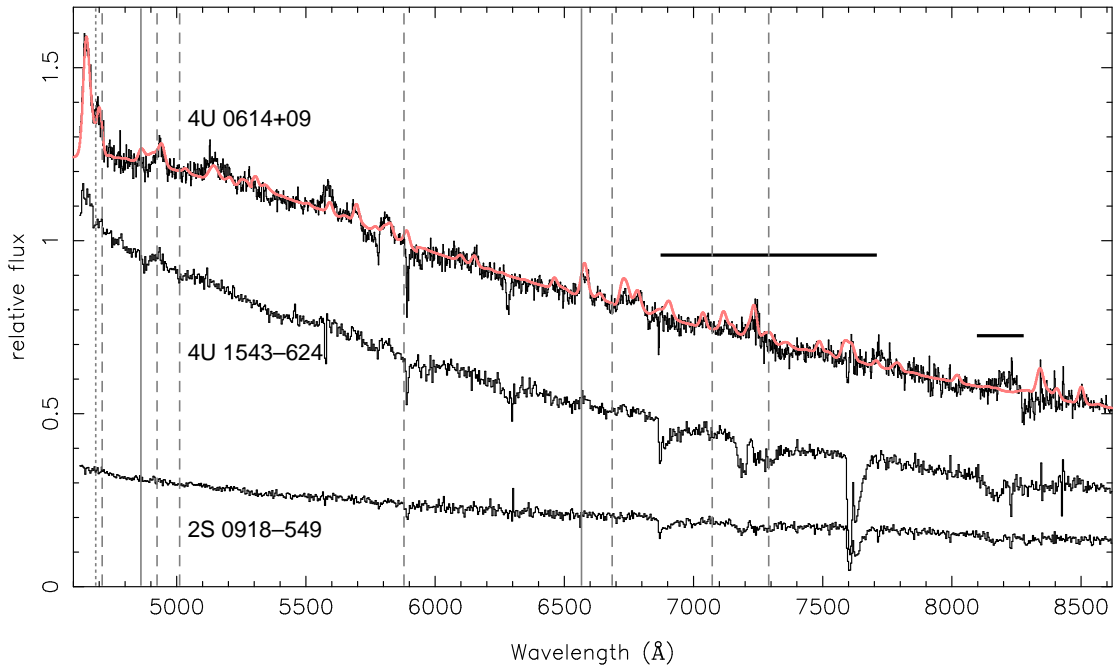}
  \caption{VLT spectra of 4U 0614+09, 4U 1543-624 and 2S 0918-549,
  showing lines from a carbon-oxygen accretion disc. From
  \citet{njm+04}}
\label{fig:2003}
\end{figure}

The results of our 2003 programme are published in \citet{njm+04} and
are summarized in Fig.~\ref{fig:2003}. We identified the features in
the spectrum of 4U 0614+09 as relatively low ionization states of
carbon and oxygen. This clearly identifies this system as an UCXB and
suggests the donor in this system is a carbon-oxygen white dwarf. The
similarity of the spectrum of 4U 1543-624 suggests it is a similar
system, while for 2S 0918-549 the spectrum didn't have a high enough
S/N ratio to draw firm conclusions, but is also is consistent with
being a similar system (and clearly does not show the characteristic
strong hydrogen emission lines of low-mass X-ray binaries). We
therefore concluded that all these systems are UCXBs.

\subsection{The known UCXB 4U 1626-67 compared to 4U 0614-09}

\begin{figure}
  \includegraphics[angle=-90,width=0.7\textwidth]{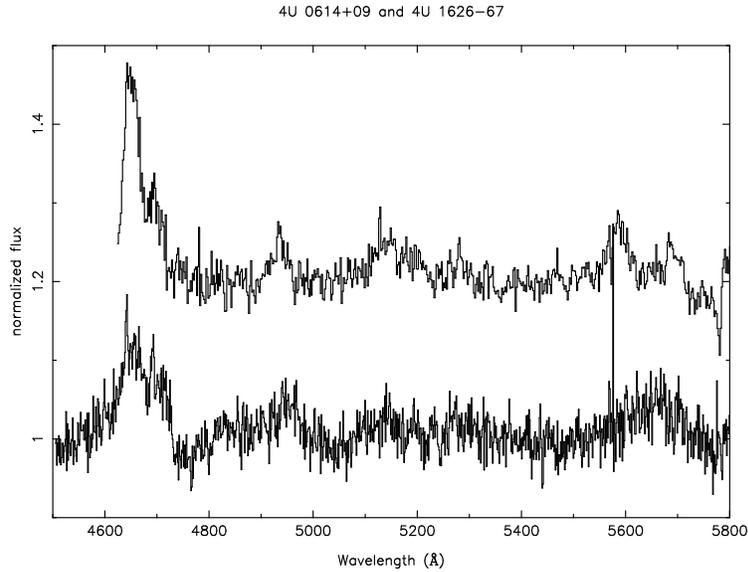}
  \caption{VLT spectra of Comparison of the carbon-oxygen spectrum of
  4U 0614+09 with the spectrum of the 42 min system 4U 1626-67.}
\label{figL0614_1626}
\end{figure}

In Fig.~\ref{figL0614_1626} we compare the blue part of our 4U 0614+09
spectrum with the VLT spectrum of the 42 min. binary 4U 1626-67, which
harbours a 7 sec X-ray pulsar \citep{mmn81}. The similarities are
remarkable. This is interesting, as strong line emission in the X-ray
spectrum of 4U 1626-67 has been identified with O and Ne lines
\citep{sch+01}.

\subsection{The preliminary 2004 results}

The results of the 2004 programme will be published in a forthcoming
paper, but we will give some preliminary results here.

\subsubsection{4U 1822-00}

Due to the faintness of 4U 1822-00 its spectrum is of low
quality. However, just as with 2S 0918-549, the spectrum does not show
hydrogen or helium lines, making it clearly different from the spectra
of hydrogen rich systems. Provisionally we classify this system as an
UCXB.

\subsubsection{4U 1556-60}

\begin{figure}
  \includegraphics[angle=-90,width=0.8\textwidth]{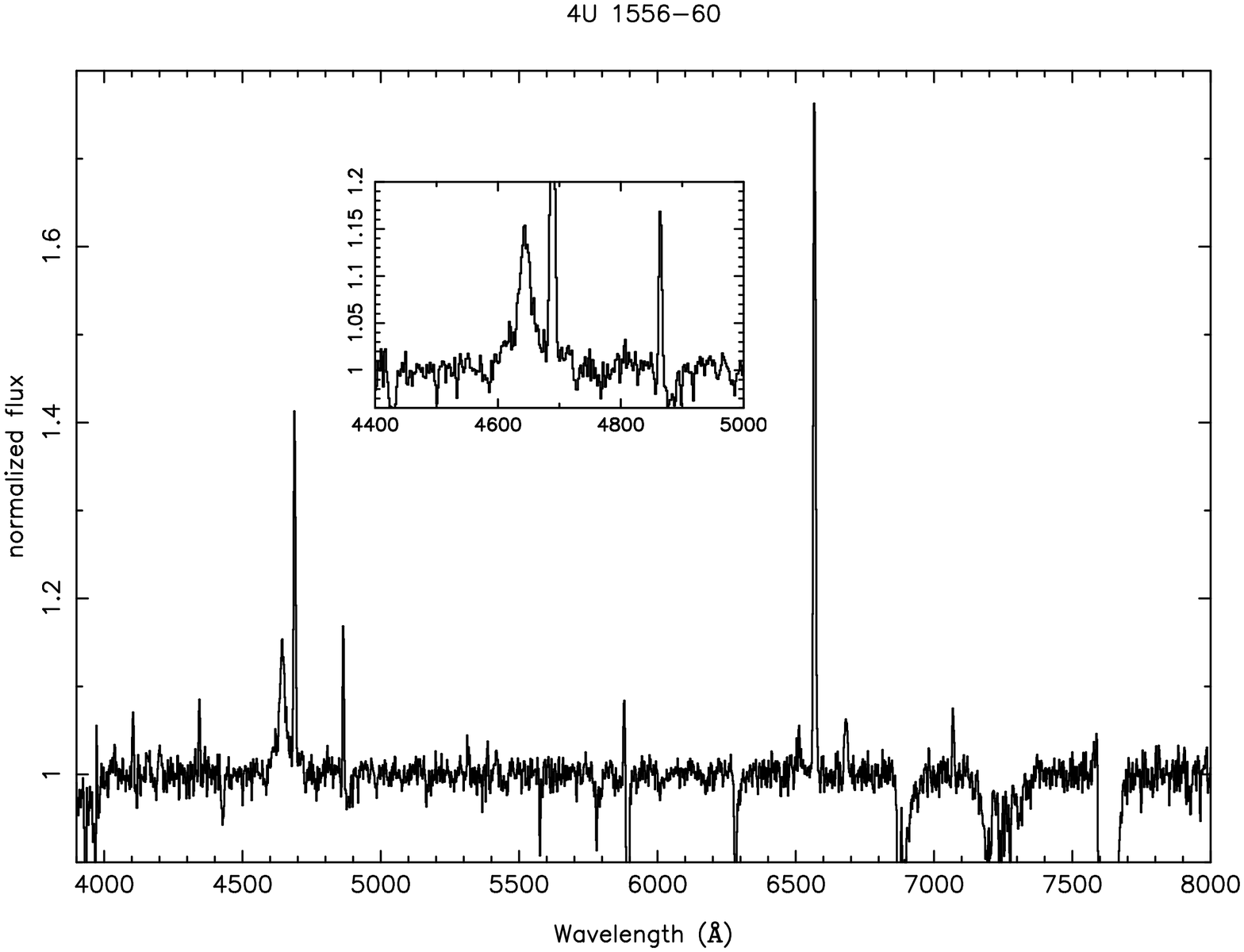}
  \caption{Normalized spectrum of 4U 1556-60, showing strong H and
  HeII emission, plus the Bowen blend at 4640\AA.}
\label{1556}
\end{figure}

Based on its ``Ne feature'' 4U 1556-60 is a good UCXB
candidate. However, its optical spectrum (Fig.~\ref{1556}) shows a
classical low-mass X-ray binary spectrum with strong Balmer lines
(4101, 4340, 4961 and 6563 \AA) and lines from HeII (4686 (very
strong), 5411 and 6678 \AA). There is also strong emission at the
Bowen blend, a C and N complex around 4640\AA\ that is driven by He
fluorescence. This system thus probably is not an UCXB, suggesting that
the ``Ne feature'' is not a unique property of UCXBs.

\subsubsection{XB 1905+00}


The spectrum of XB 1905+00 shows the standard features of an early G
star. This puzzling result is possibly due to a chance alignment. The
acquisition image of the object obtained with a seeing of 0.6 arcsec
suggest the source actually is a blend of two objects. In that case
the optical counterpart of XB 1905+00 would be the fainter of the two
stars.  Note that this system was in quiescence at the time of our
observation, making the counterpart much fainter than when it was
found by \citet{ci85}.

\subsubsection{The known UCXB XB 1916-05}

\begin{figure}
  \includegraphics[angle=-90,width=0.7\textwidth]{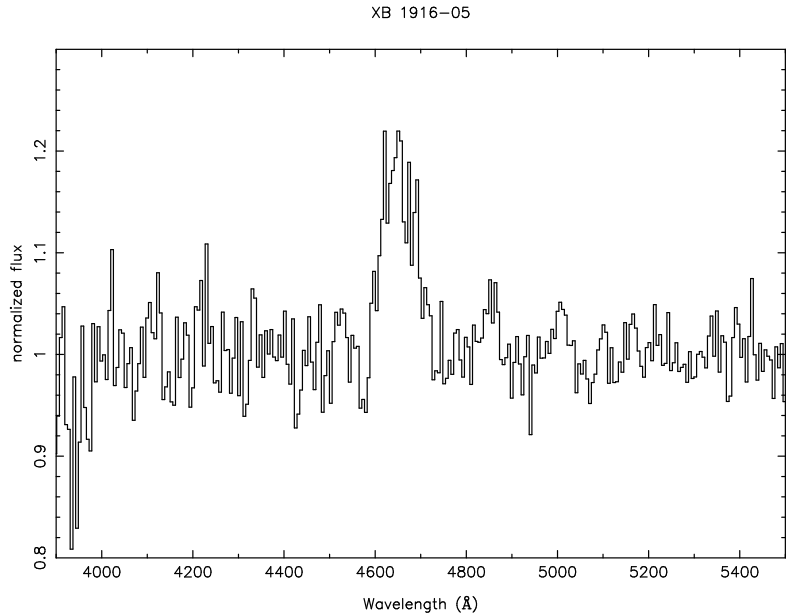}
  \caption{Normalized (and rebinned) spectrum of the known UCXB XB
  1916-05, showing broad features, like 4U 0614, but possibly also
  hints of He and H emission.}
\label{fig:1916}
\end{figure}

In Fig.~\ref{fig:1916} we show the spectrum of the 50 min UCXB XB
1916-05. Again it shows some features that are very similar to the
spectra of 4U 0614+09 and 4U 1626-67, but in addition broad emission
around 4540\AA. A possible origin could be He, as there is a HeII line
at 4541\AA. Some of the other features also coincide roughly with
positions of HeII lines, but these are very close (within a few \AA)
of the Balmer lines, making identification very difficult.

\section{Conclusions}

We have started a systematic study of the optical spectra of known and
candidate UCXBs. The first results are both interesting and promising:
we have confirmed the ultra-compact nature of many of the candidates
and have uncovered a variety of optical spectra, from pure
carbon-oxygen spectra to spectra probably still showing signs of
helium (and possibly hydrogen). One candidate (4U 1556-60) shows a
classical low-mass X-ray binary spectrum, making it unlikely that this
system is an UCXB. The faintness of these systems makes spectroscopic
period determination almost impossible, so that periods will have to
be found photometrically, as was recently done with 4U 1543-624 \citep{wc04}.






\bibliographystyle{aipproc}   

\bibliography{binaries.bib,journals.bib}

\begin{thebibliography}{13}
\expandafter\ifx\csname natexlab\endcsname\relax\def\natexlab#1{#1}\fi
\providecommand{\enquote}[1]{``#1''}
\expandafter\ifx\csname url\endcsname\relax
  \def\url#1{\texttt{#1}}\fi
\expandafter\ifx\csname urlprefix\endcsname\relax\def\urlprefix{URL }\fi
\providecommand{\eprint}[2][]{\url{#2}}

\bibitem[{Nelson} et~al.(1986)]{nrj86}
L.~A. {Nelson}, S.~A. {Rappaport}, and P.~C. {Joss}, \emph{\apj}, \textbf{304},
  231--240 (1986).

\bibitem[{Markwardt} et~al.(2002)]{mss+02}
C.~B. {Markwardt}, J.~H. {Swank}, T.~E. {Strohmayer}, J.~J.~M.~i. {Zand}, and
  F.~E. {Marshall}, \emph{\apjl}, \textbf{575}, L21--L24 (2002).

\bibitem[{Remillard} et~al.(2002)]{rss02}
R.~A. {Remillard}, J.~{Swank}, and T.~{Strohmayer}, \emph{\iaucirc},
  \textbf{7893}, 1 (2002).

\bibitem[{Markwardt} et~al.(2003)]{mjs03}
C.~B. {Markwardt}, M.~{Juda}, and J.~H. {Swank}, \emph{\iaucirc},
  \textbf{8095}, 2 (2003).

\bibitem[{Podsiadlowski} et~al.(2002)]{prp02}
P.~{Podsiadlowski}, S.~{Rappaport}, and E.~D. {Pfahl}, \emph{\apj},
  \textbf{565}, 1107--1133 (2002).

\bibitem[{Van Paradijs} and {McClintock}(1994)]{vm94}
J.~{van Paradijs}, and J.~E. {McClintock}, \emph{\aap}, \textbf{290}, 133--136
  (1994).

\bibitem[{Wang} and {Chakrabarty}(2004)]{wc04}
Z.~{Wang}, and D.~{Chakrabarty}, \emph{\apjl}, \textbf{616}, L139--L142 (2004).

\bibitem[Juett et~al.(2001)]{jpc00}
A.~M. Juett, D.~Psaltis, and D.~Chakrabarty, \emph{\apjl}, \textbf{560},
  L59--L63 (2001).

\bibitem[{Bildsten}(1995)]{bil95}
L.~{Bildsten}, \emph{\apj}, \textbf{438}, 852--875 (1995).

\bibitem[Nelemans et~al.(2004)]{njm+04}
G.~Nelemans, P.~G. Jonker, T.~R. Marsh, and M.~van~der Klis, \emph{\mnras},
  \textbf{348}, L7 (2004).

\bibitem[{Middleditch} et~al.(1981)]{mmn81}
J.~{Middleditch}, K.~O. {Mason}, J.~E. {Nelson}, and N.~E. {White},
  \emph{\apj}, \textbf{244}, 1001--1021 (1981).

\bibitem[Schulz et~al.(2001)]{sch+01}
N.~S. Schulz, D.~Chakrabarty, H.~L. Marshall, C.~R. Canizares, L.~C. Lee, and
  H.~J., \emph{\apj}, \textbf{563}, 941 (2001).

\bibitem[{Chevalier} and {Ilovaisky}(1985)]{ci85}
C.~{Chevalier}, and S.~A. {Ilovaisky}, \emph{Space Science Reviews},
  \textbf{40}, 443 (1985).

\end{thebibliography}

\IfFileExists{\jobname.bbl}{}
 {\typeout{}
  \typeout{******************************************}
  \typeout{** Please run "bibtex \jobname" to optain}
  \typeout{** the bibliography and then re-run LaTeX}
  \typeout{** twice to fix the references!}
  \typeout{******************************************}
  \typeout{}
 }

\end{document}